\documentclass[a4paper,prb,twocolumn,superscriptaddress,showpacs]{revtex4-1}
\usepackage{amssymb}
\usepackage{graphicx}
\usepackage{epstopdf}
\usepackage{amsmath}
\usepackage{natbib}
\usepackage{hyperref}

\begin{document}

\title{Desiging Artificial Lieb Lattice on Metal Surface}
\author{Wen-Xuan Qiu}
\affiliation{School of Physics and Wuhan National High Magnetic field center,
Huazhong University of Science and Technology, Wuhan 430074,  China}
\author{Shuai Li}
\affiliation{School of Physics and Wuhan National High Magnetic field center,
Huazhong University of Science and Technology, Wuhan 430074,  China}
\author{Jin-Hua Gao}
\email{jinhua@hust.edu.cn}
\affiliation{School of Physics and Wuhan National High Magnetic field center,
Huazhong University of Science and Technology, Wuhan 430074,  China}

\author{Yi Zhou}
\affiliation{Department of Physics, Zhejiang University, Hangzhou, China}
\affiliation{Collaborative Innovation Center of Advanced Microstructures, Nanjing 210093,
China}

\author{Fu-Chun Zhang}
\affiliation{Department of Physics, Zhejiang University, Hangzhou, China}
\affiliation{Collaborative Innovation Center of Advanced Microstructures, Nanjing 210093,
China}

\begin{abstract}
 Recently, several experiments \cite{nature2012,lin2014} have illustrated that metal surface electrons can be manipulated to form a two dimensional (2D) lattice by depositing a designer molecule lattice on metal surface.  This offers a promising new technique to construct artificial 2D electron lattices. Here we theoretically propose a molecule lattice pattern to realize an artificial Lieb lattice on metal surface, which shows a flat electronic band due to the lattice geometry.   We show that the localization of electrons in the flat band  may be understood from the viewpoint of  electron interference, which may be probed by measuring the local density of states with the scanning tunnelling microscopy.  Our proposal may be readily implemented in experiment and may offer an ideal solid state platform  to investigate the novel flat band physics of the Lieb lattice.
\end{abstract}

\maketitle
The two-dimensional (2D) lattices with flat electron bands are of special research interests\citep{tasaki_nagaokas_1998,wuyongshi2014,zhaoliu2013}. The electrons in such flat bands are localized due to the destructive wave interference resulted from the special lattice geometries. By the flat band, we  mean the kinetic energy of electrons to be quenched, so that tiny interactions can induce various exotic many-body states, such as ferromagnetism\cite{tasaki_nagaokas_1998, lieb98,*mielke91,*shen1994}, Wigner crystal\cite{wuc2007} and superconductivity\cite{scflat07,sc2016,kopnin2011}. More recently, it has been pointed out that if nontrivial topology can be introduced into flat band, fractional quantum Hall states may occur in the absence of an external magnetic field\cite{wen2011,neupert2011,sunkai2011,shengdn2011,bernevig}. Due to these novel properties, great efforts have been made  in the recent years to  search 2D flat band systems in real materials\cite{wangzf2013} as well as in artificial 2D lattice systems such as cold atom \cite{xing2010,goldman2011,apaja2010}, photonic crystal\cite{prl032015,prl042015}, quantum dot lattice\cite{qd2002} and circuit QED lattice\cite{huyong2016}.

\begin{figure}
\centering
\includegraphics[width=7cm]{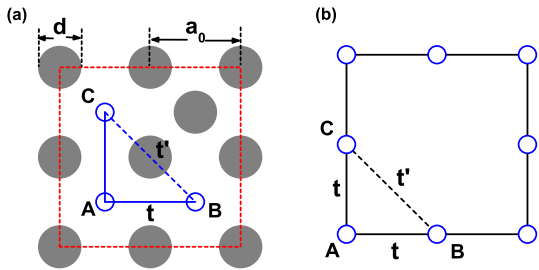}
\caption{(Color online) (a) The muffin-tin potential proposed in this paper to induce artificial Lieb lattice with nearly flat electron band.  In the model calculation, the potential is $\textrm{U}_0$ inside the gray dots which represent the absorbed molecules and zero outside the dots. $d$ is the diameter of the potential dots and $a_0$ is the lattice constant of the molecule lattice.  The red dashed lines indicate the unit cell of the 2D system in the presence of the muffin-tin potential.  The blue circles represent the three sites (A, B, C) in a unit cell of the artificial Lieb lattice for the surface electrons. $t$ and $t'$ are the NN and NNN hopping integrals, respectively. (b) The equivalent Lieb lattice.
}
\label{muffintin}
\end{figure}

The Lieb lattice is one of  the most well-known 2D flat band lattices. It is a line centered square lattice, and is composed of three atoms (A,B,C) in one unit cell (see in Fig.  \ref{muffintin}).  The dispersion of the tight-binding model with nearest neighbor hopping on the Lieb lattice consists of three bands including  a flat band in the middle of the spectra. At half filling,  infinitesimal on-site Coulomb interaction can induce a ferromagnetic ground state, i.e. the flat band ferromagnetism\cite{lieb98,shen1994}. More interestingly, away from the half filled, the spin-orbit coupling can induce various exotic topologically nontrivial phases in the Lieb lattice\cite{MFranz2010,zhaoan2012,yaohong2015}. Under certain conditions, the flat band of the Lieb lattice may have a nonzero Chern number, which is crucial to realize fractional Chern insulator\cite{fci2015,chern2015,*integer2016}.  Recently, it is reported that the Lieb lattice has been realized in photonic crystal\cite{prl032015, prl042015} and cold atom system\cite{science2015}. However, searching the Lieb lattice of electrons in a solid state system remains a challenge and is badly needed for the intriguing theoretical proposals.

In this work, we propose a  practical scheme to induce a tunable artificial Lieb lattice on metal surface. The basic idea is motivated by a recent  experiment where an artificial graphene is successfully realized on metal surface\cite{nature2012,lin2014}. In that experiment, CO molecules are assembled on Cu(111) surface by scanning probe technique to form a hexagonal lattice, in which each CO molecule becomes a repulsive potential center on the metal surface and the surface electrons are effectively confined on the discrete sites of an artificial  honeycomb lattice and have a linear dispersion\cite{nature2012,nanoscale2016,popodft2014}.  Here we theoretically propose a special arrangement of the CO molecules and  show that the metal surface electrons can be transformed into a Lieb lattice with a flat or nearly flat band.   We first use numerical method to find the electronic band structure of the metal surface electrons when the proposed CO molecules are absorbed, followed by a tight binding model fitting with nearest neighbor (NN) and next nearest neighbor (NNN) hopping integrals on the Lieb lattice.  Due to a small but finite NNN hopping, the flat band  on this artificial Lieb lattice is bended with a narrow band width and a large DOS, and small interactions may lead to instabilities of the long range ordered states. The flatness of the flat band electrons on this artificial Lieb lattice depends on the strength of the repulsive potential applied by the absorbed molecule.  The larger the repulsive potential is, the flatter the band is. Meanwhile, the filling number of electrons on the Lieb lattice is also controllable.
  %Since this artificial Lieb lattice is on metal surface,  its electronic properties, e.g. the LDOS, spin order and edge mode, can be readily detected by the scanning tunnelling microscopy (STM) measurement.%
 An interesting issue is the wave function localization of the flat band electrons.  Due to the geometric phase cancellation,  the flat band electrons of the Lieb lattice are localized only at the B and C sites, which gives a unique LDOS pattern for the flat band electrons. This special type of LDOS is demonstrated in our numerical simulation, which can  be directly probed in the scanning tunnelling microscopy (STM).

Our proposal is of several merits: (1) it is a solid state electronic system, and has intrinsic Coulomb interaction which is essential for these exotic many-body states of flat band fermions; (2) because that the artificial Lieb lattice is made on the metal surface, we may apply various techniques available in condensed matter physics to detect the electronic structure of the flat band, e.g. the STM to detect the LDOS, possible charge order (or spin order), and even the edge states which is crucial for the topological nontrivial phases.
%including the STM for LDOS as well as the edge states and the angle resolved photoemission spectroscopy for dispersions and thermodynamic measurement for the possible long range ordering.%
(3) It should be accessible to implement this scheme in experiment since that the same technique has been used to realize an artificial honeycomb lattice (molecule graphene).

In order to make our proposal more realistic, we consider the system of Cu(111) surface and CO molecules  which was used to construct the molecule graphene. The Cu surface state can be viewed as a 2D electron gas (2DEG) with parabolic dispersion, and the Hamiltonian is $H_{\textrm{Cu}}=\frac{\hbar^2 k^2}{2m^*}$, where $m^* = 0.38 m_e$ is the effective mass. The Fermi energy is $E_f = 0.45$ eV from the band bottom. When a CO molecule is absorbed on the Cu surface, it exerts a repulsive potential on the underlying surface electrons.  Thus, if the CO molecules are assembled into a lattice on metal surface, the surface electrons feel a lateral periodic potential  and can be forced into a  fermionic lattice system. Note that, this Cu/CO system is a quantum antidot system since the CO molecule will deplete the underlying surface electrons. In theory, the potential of the CO molecule lattice can be approximated as a muffin-tin potential $U(r)$. We can obtain the energy dispersion of the modified Cu surface states via solving the system Hamiltonian $H=H_{\textrm{Cu}}+U(r)$ by plane wave method\cite{nanoletter2009,nanoscale2016}.

The muffin-tin potential we propose to induce an artificial electron system on a Lieb lattice is illustrated in Fig. \ref{muffintin} (a).  $\textrm{U}_0$ ($>0$) inside the gray dots and zero elsewhere. Here a gray dot represents a CO molecule or the potential of CO molecule. Note that a Lieb lattice can be achieved via removing one quarter of the sites of the square lattice (see Fig. 1(b)).  Following this basic idea, we first arrange the CO molecules  into a square lattice, where lattice constant is $a_0$ as shown in Fig. \ref{muffintin} (a). Consequently, due to the repulsive periodic potential of the CO molecule lattice,  the metal surface electrons are forced also into a square lattice with the same lattice constant, where the lattice sites (blue circles) are at the center of the squares formed by four neighbouring molecules. Then, we delete one of the four sites of surface electron lattice by positioning an additional molecule to the center of the upper right square, as shown in Fig. 1(a). Finally, we get an artificial Lieb lattice of surface electrons in such antidot lattice system.
The red dotted line in Fig. \ref{muffintin} (a) gives the unit cell of the muffin-tin potential, and the corresponding Lieb lattice of the surface electrons is  shown in Fig. \ref{muffintin} (b). We use $t$ and $t'$ to denote the NN and NNN hopping integrals, respectively.

Now, we  show that the low energy physics of the artificial 2D system is equivalent to electrons on a Lieb lattice. We first calculate the energy dispersion  of the surface electrons in the presence of the molecules by the plane wave method, and then we show that the three lowest mini-bands, induced by the lateral periodic potential, can be well described by the tight binding model of the Lieb lattice.

%As mentioned above, due to the geometric phase cancellation, the wave function of the flat band in Lieb lattice has %finite amplitude  only in  the B and C sites, which results in a special LDOS pattern.  Our numerical calculations %clearly illustrate this characteristic of flat band.

Let us first discuss the calculation of energy dispersion.  With the plane wave method, there are three parameters of the muffin-tin potential in our model: the lattice constant $a_0$, the potential value $U_0$ and the potential diameter $d$ (see in Fig. \ref{muffintin}). Here, $a_0$ is the distance between adjacent molecules, which is tunable in experiment.  $U_0 = 7$ eV and $d=0.5$ nm are the reasonable values of Cu/CO system got by fitting the experiments\cite{nanoscale2016}.
%In the numerical calculation, for the muffin-tin potential, there are three parameters as shown in Fig. \ref{muffintin} (a), i.e. the potential value $\textrm{U}_0$, the diameter of the potential %disk $d$, and the molecule lattice distance $a_0$.
%By fitting the experimental results of artificial graphene system, for the Cu(111)/CO system, we suggest that the potential value $\textrm{U}_0$ of the CO molecule should be of the order of several %eV, and here we use $\textrm{U}_0 = 7$ eV which can give  a good approximation to the experiment. We set $d=0.5$ nm for the potential disk diameter.  Note that $\textrm{U}_0$ and $d$  can be changed %by choosing different experimental systems, e.g. using Au or other kinds of molecule. The molecule lattice distance $a_0$ is a tunable parameter in experiment, which can determine not only the band %dispersion but also the position of the Fermi level. In calculation, we assume that the absorbed molecules do not change the total number of the surface electrons.

Meanwhile, the tight binding Hamiltonian of the Lieb lattice is
\begin{equation}
H=- \sum_{ij,\sigma} t_{ij}c^+_{i\sigma}c_{j\sigma} + \sum_{i\sigma} \epsilon_i c^+_{i\sigma}c_{i\sigma} ,
\end{equation}
where $c^+_{i\sigma}$ creates an electron with spin $\sigma$ on lattice site $i$.  The unit cell of the Lieb lattice is given by A, B, C sites shown in Fig. \ref{muffintin}, and we consider the NN hopping $t$ and NNN hopping $t'$.  For square lattice, there is an approximate relation about the NN hopping $t$ with the underlying 2D system,
\begin{equation}\label{nnhopping}
t = \frac{\hbar^2}{2m^* a_0^2},
\end{equation}
 which we will choose for  the Lieb lattice model with an additional fitting parameter of the NNN hopping $t'$.  Note that the circumstance of the B (and C) site is different from that of A site in Fig. \ref{muffintin} (a), so that in the tight binding model the onsite energy of B and C sites are equal, but can be different from that of A site. The Hamiltonian is a $3 \times 3$ matrix
\begin{widetext}
\begin{equation}
H(k)=\left( \begin{array}{ccc}
\epsilon_A& -2t\cos(k_xa_0)& -2t\cos(k_ya_0)   \\
-2t\cos(k_xa_0)&\epsilon_B& -4t' \cos(k_xa_0) \cos(k_ya_0)   \\
-2t\cos(k_ya_0)&  -4t' \cos(k_xa_0) \cos(k_ya_0) &\epsilon_C\\
\end{array}
\right)
\end{equation}
\end{widetext}
 We set $\epsilon_B = \epsilon_C =\epsilon_0 +\Delta$ and $\epsilon_A = \epsilon_0 - \Delta$, where $\epsilon_0$ and $\Delta$ are parameters to be determined. When $t'=0$, the dispersion of the tight binding model is simple, and it gives two dispersive  bands $\epsilon_{\pm} (k)= \epsilon_0 \pm \sqrt{4t^2 [\cos^2 (k_x a_0) + \cos^2(k_y a_0)]+\Delta^2}$ and one dispersionless  flat band $\epsilon(k)=\epsilon_0 + \Delta$.   It should be noted that, when $\Delta=0$, i.e. $\epsilon_A = \epsilon_B =\epsilon_C$, the three bands touch at the M point in the FBZ, while if $\Delta \neq 0$, the flat band touch only one dispersive band at M point and a gap about $2|\Delta|$ can be observed at M point. In addition, it can be seen from the TB model that the wave function of the flat band electrons of  Lieb lattice  only  locates  at B and C sites.
 If $t'$ is nonzero, the analytical expressions of the energy dispersion become more complex and numerical calculation may be a better way.
 However, the properties above approximately  hold when $t'$ is not too large.

\begin{figure}[!hb]
\centering
\includegraphics[width=8.5cm]{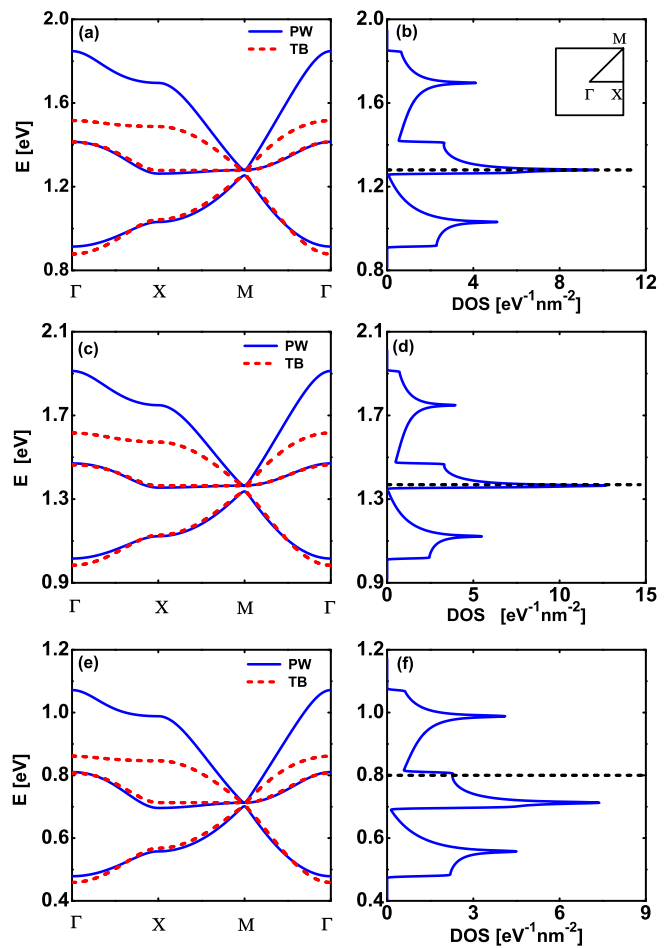}
\caption{(Color online) Left panels:  three lowest bands (blue lines) calculated within plane wave (PW) method for the muffin-tin potential given in Fig. \ref{muffintin} and the fitting electron bands (red dotted lines) within the tight-binding (TB) model of the Lieb lattice in Eq. (1) for 3 sets of the parameters in Fig. \ref{muffintin}  (a). $\textrm{U}_0 = 7$ eV, $a_0=0.95$ nm and $d=0.5$ nm.  The fitting parameters in TB are $t=0.11$ eV, $t'=0.034$ eV, $\Delta=0.012$ eV and $\epsilon_0=1.27$ eV. (c) $\textrm{U}_0 = 9$ eV, $a_0=0.95$ nm and $d=0.5$ nm. The fitting parameters in Eq. (1) are $t=0.11$ eV, $t'=0.025$ eV, $\Delta=0.014$ eV and $\epsilon_0=1.35$ eV.  (e)  $\textrm{U}_0 = 7$ eV, $a_0=1.2$ nm and $d=0.5$ nm. The fitting parameters in Eq. (1) are  $t=0.069$ eV, $t'=0.023$ eV, $\Delta=0.006$ eV and $\epsilon_0=0.71$ eV. Right panel:  (b),(d), (f) are the corresponding DOS for the cases of (a), (c),(e), respectively. The black dotted line is the Fermi level.
}
\label{band}
\end{figure}

\begin{figure}
\centering
\includegraphics[width=7cm]{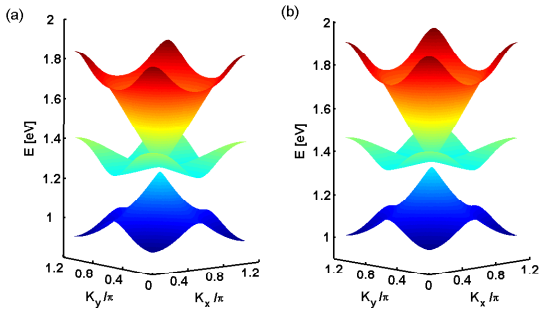}
\caption{(Color online) The lowest three energy bands calculated by plane wave method in the FBZ. (a) is for the case of Fig. \ref{band} (a). (b) is for the case of Fig. \ref{band} (c).
}
\label{3D}
\end{figure}

 The numerical results of the energy bands are shown in Fig. \ref{band}. In Fig. \ref{band} (a), we plot the lowest three bands of the calculated band structure got by the plane wave method, with the lattice constant $a_0=9.5$ $\AA$ (blue solid lines).  Meanwhile, the tight binding bands of the Lieb lattice are  given as a comparison (red dashed lines), with a $t$ got by Eq. \ref{nnhopping} and a fitted $t'$. We see that, the three plane wave bands, especially the lowest two, can be well described by the tight binding model of Lieb lattice. Note that here, once $a_0$ is given, the NN hopping $t$ is fixed and $t'$ is got by fitting the band shape, while $\epsilon_0$ and $\Delta$ are got by fitting the energy gap at M point.
 One important characteristic is that, due to the nonzero $t'$, the middle band of the Lieb lattice is not  completely flat  but  bended. However, it still has a narrow band width and a giant DOS as given in Fig. \ref{band} (b), which is much larger than that near  Van Hove singularities.  In addition, because that the onsite energy of B (and C) site is different from that of A site, i.e. $\Delta \neq 0$, the flat band only touches  the upper band at M point in FBZ, and has a small gap about $2|\Delta| \approx 24$ meV at M point from the lower band. Note that, since that the  middle band is bended away from the M point, this small gap  can not be observed from the DOS.   This is an intrinsic property of this artificial Lieb lattice. We then plot the calculated bands in the whole FBZ in Fig. \ref{3D} (a). All the characteristics mentioned above  are also shown clearly.
 %The numerical results of the band dispersion is our first proof that this scheme can produce an artificial Lieb lattice on metal surface.

In this artificial Lieb lattice, the flatness of the flat band is tunable.
Actually, the flatness of the middle band of Lieb lattice depends on the NNN hopping $t'$. Smaller $t'$ gives a flatter band.  In this artificial Lieb lattice, $t'$  is determined by the value of muffin-tin potential $\textrm{U}_0$. This point can be understood from  Fig. \ref{muffintin} (a). We see that $t'$ is the hopping between B and C sites, which is blocked by an in-between molecule. Intuitively, the stronger the potential $\textrm{U}_0$ is the smaller $t'$ is, and thus a flatter band can be achieved.  This understanding is confirmed by the numerical results with a larger value of  $\textrm{U}_0$,  which are plotted in Fig. \ref{band} (c), (d) and Fig. \ref{3D} (b). With $\textrm{U}_0 = 7$ eV in Fig. \ref{band} (a), the band width of the middle band is about 150 meV, while with $\textrm{U}_0 = 9$ eV it becomes 120 meV in Fig. \ref{band} (c).  So, a flatter band can be achieved by choosing a proper molecule with stronger repulsive potential.

Meanwhile, the position of the Fermi level is also controllable in this system by tuning the lattice constant $a_0$.  Note that, in order to investigate the physics of flat band,  the Fermi level should be in or close to the flat band. Or equivalently,  the filling of Lieb lattice should be in the region between $\frac{1}{3}$ and $\frac{2}{3}$. An  assumption here is that the total electron number of the metal surface states is fixed even in the presence of the absorbed molecules. Therefore, with different $a_0$, the band structure is different and thus the position of Fermi level is changed accordingly. We now give some quantitative estimations. The filling of this artificial Lieb lattice is  $\frac{2}{3}N_e a_0^2$, where $N_e$ is the electron density of the metal surface. For Cu(111) surface,  to access the flat band, $a_0$ should be in the region $0.8 \ \textrm{nm} < a_0 < 1.2 \ \textrm{nm}$, where $N_e$ is about $0.72$ $\textrm{nm}^{-2}$. For  half filling which has an exact ferromagnetic ground state, $a_0$ is about $1.0$ nm.   For larger $a_0$, the Fermi level is at the upper band, as shown in Fig. \ref{band} (e) and (f). These estimations are general for the Cu(111) surface no matter which kind of molecule is used in experiment. We emphasize that the required region of $a_0$ above has already been achieved in the molecule graphene experiments\cite{nature2012,lin2014}.

%Quantitatively,  the Fermi level is determined  by solving the number equation
%\begin{equation}
%N_e = \int^{\mu}_{E_0} d\epsilon \rho(\epsilon),
%\end{equation}
%where $\mu$ is the chemical potential, $E_0$ is the band bottom, $\rho(\epsilon)$ is the DOS of the artificial Lieb lattice.  The electron density of the metal surface is a known number, %$N_e=\rho_{2D}\times E_f$. For example, $N_e$ for the Cu(111) surface is about $0.72$ $\textrm{nm}^{-2}$. In Fig. \ref{band} (b) and (d), we plot the calculated position of Fermi level (black dashed %line).  It is shown that, by carefully choosing the  parameters
%of muffin-tin potential (i.e. $a_0$ and $\textrm{U}_0$), the Fermi level can be tuned into the flat band region. A better way is to estimate the filling of the artificial Lieb lattice. The total %number of electron in one unit cell is $N_e \times 4a_{0}^2$. Considering that there are three sites in one unit cell, the filling of the artificial Lieb lattice is  $\frac{2}{3}N_e a_0^2$. Thus, to %access the flat band, the lattice constant $a_0$ should be in the region $0.8 \ \textrm{nm} < a_0 < 1.2 \ \textrm{nm}$.  For the case of half filling which has an exact ferromagnetic ground state, %$a_0$ is about $1.0$ nm.  This estimation is general for the Cu(111) surface no matter which kind of molecule is used in experiment.  For larger $a_0$, the Fermi level is at the upper band, as shown %in Fig. \ref{band} (e) and (f). Note that $a_0$ can not be too small, otherwise the muffin-tin potential will overlap.
\begin{figure}
\centering
\includegraphics[width=8cm]{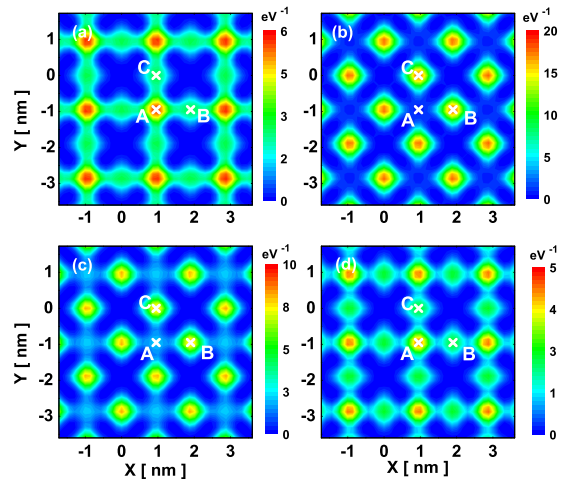}
\caption{(Color online) The LDOS pattern in real space with different energy. The potential parameters are the same as that used in Fig. \ref{band} (c).   (a) $\epsilon=1.2$ eV. (b)  $\epsilon=1.37$ eV. (c) $\epsilon=1.5$ eV. (d) $\epsilon=1.6$ eV
}
\label{LDOS}
\end{figure}

The second illustration for the Lieb lattice is the special LDOS pattern of the flat band electrons. As described above, one intriguing property of the flat band of Lieb lattice is that the wave function is localized on the B and C sublattices, and vanish on the A sublattice. We reveal this property by calculating the LDOS of surface electron
\begin{equation}
\textrm{LDOS}(r,\epsilon) = \sum_{nk\sigma} |\Phi_{nk\sigma}(r)|^2 \delta(\epsilon-\epsilon_{nk})
\end{equation}
which can be directly measured by the STM.
From the definition of the LDOS, it is easy to see that, if it is a Lieb lattice, the LDOS of the flat band electrons should have finite values around the B and C sublattices and be zero around the A sublattice.
 In Fig. \ref{LDOS}, we plot the LDOS of the artificial Lieb lattice with different energy, where the corresponding bands are given in Fig. \ref{band} (c).  In Fig. \ref{LDOS} (a), $\epsilon=1.2$ eV and thus the electrons are from the lower band. We see that  electron distribution is nonzero around all the three sites , but the amplitude around the A site is much larger than that around the B (and C) site. This is consistent with the TB results. The LDOS with $\epsilon=1.37$ eV is given in Fig. \ref{LDOS} (b), where the electrons mainly are from the middle band (flat band). Obviously, the LDOS becomes nearly zero around A sublattice and is finite around B and C sublattices. That is just the wave function localization phenomenon of Lieb lattice.
 If slightly increasing the energy, at $\epsilon=1.5$ eV, there are electrons from both the middle band and upper band. Now the electron distribution at A site becomes nonzero but tiny, while the electrons are still mainly distributed at B and C sites (see in Fig. \ref{LDOS} (c)). It implies that in this case the electrons from the flat band dominate. Increasing the energy further, we plot the LDOS pattern with $\epsilon=1.6$ eV in Fig. \ref{LDOS} (d), where the electrons are only from the upper band. The case is the similar as that of  lower band. So, in the STM experiment, the LDOS for different energy (i.e. different bias voltage) can give an direct evidence of the wave function localization of flat band electrons, and thus can be used to identify the flat band states as well as the Lieb lattice. We would like to emphasize that, in this system, the wave function localization of flat band is an interference phenomenon of the surface electrons in the presence of a special absorbed molecule lattice. In fact, the absorbed molecule (or atom) induced interference phenomena  are  usual on metal surface, e.g. standing wave of surface electron and the celebrated quantum corral. Thus, it should not be surprising that this wave function localization can be observed in experiment.

In summary, we theoretically propose a scheme to realize an artificial Lieb lattice on the metal surface via the technique reported in the recent molecule graphene experiments. All the requirements of this scheme have already been fulfilled in the recent molecule graphene experiments, so that it can be easily implemented in experiment. Considering that the external magnetic field, as well as the spin-orbit coupling, can be readily induced, we think that this artificial Lieb lattice on metal surface is an ideal solid state platform to study the novel flat band physics of Lieb lattice, e.g. the wave function localization of the flat band electrons, the correlated states and the possible topological nontrivial phases. We point out that the  LDOS measured by STM can be  used to identify the wave function localization of the flat band electrons.

 This work is supported by the National Science Foundation of China (Grants No. 11274129, No. 11534001, No.11374256/11274269/11674278), National Basic Research Program of China (No.2014CB921201/2014CB921203)£¬and National Key R\&D Program of the MOST of China (No.2016YFA0300202),.
F.C.Z was also supported by the Hong Kong University Grant Council via Grant No. AoE/P-04/08.

%\appendix*
%\section{The Energy Dispersion of the TB model of Lieb Lattice with NN hopping $t$ and NNN hopping $t'$}
%Here, we give the analytical experssions of the energy bands for the Hamitonian in Eq. \ref{band}, which is much more complicated than the $t'=0$ case.

%\bibliographystyle{plain}

%\bibliographystyle{apsrev4-1}
\bibliography{0625lipaperref}

\end{document}